\pdfoutput=1
\RequirePackage{ifpdf}
\ifpdf 
\documentclass[pdftex]{sigma}
\else
\documentclass{sigma}
\fi

\numberwithin{equation}{section}

\newtheorem{Theorem}{Theorem}[section]
\newtheorem{Lemma}[Theorem]{Lemma}
\newtheorem{Hyp}[Theorem]{Hypothesis}
 { \theoremstyle{definition}
\newtheorem{Definition}[Theorem]{Definition}
\newtheorem{Remark}[Theorem]{Remark} }

\begin{document}

\allowdisplaybreaks

\newcommand{\arXivNumber}{1505.06579}

\renewcommand{\thefootnote}{}

\renewcommand{\PaperNumber}{031}

\FirstPageHeading

\ShortArticleName{Tetrahedral Equation and Higher Hamiltonians of Quantum Integrable Systems}

\ArticleName{Zamolodchikov Tetrahedral Equation and Higher\\ Hamiltonians of $\boldsymbol{2d}$ Quantum Integrable Systems\footnote{This paper is a~contribution to the Special Issue on Recent Advances in Quantum Integrable Systems. The full collection is available at \href{http://www.emis.de/journals/SIGMA/RAQIS2016.html}{http://www.emis.de/journals/SIGMA/RAQIS2016.html}}}

\Author{Dmitry V.~TALALAEV}

\AuthorNameForHeading{D.V.~Talalaev}

\Address{Geometry and Topology Department, Faculty of Mechanics and Mathematics,\\ Moscow State University, Moscow, 119991 Russia}
\Email{\href{mailto:dtalalaev@yandex.ru}{dtalalaev@yandex.ru}}
\URLaddress{\url{http://higeom.math.msu.su/people/talalaev/english.htm}}

\ArticleDates{Received January 17, 2017, in f\/inal form May 13, 2017; Published online May 22, 2017}

\Abstract{The main aim of this work is to develop a method of constructing higher Hamiltonians of quantum integrable systems associated with the solution of the Zamolodchikov tetrahedral equation. As opposed to the result of V.V.~Bazhanov and S.M.~Sergeev the approach presented here is ef\/fective for generic solutions of the tetrahedral equation without spectral parameter. In a sense, this result is a~two-dimensional generalization of the method by J.-M.~Maillet. The work is a~part of the project relating the tetrahedral equation with the quasi-invariants of 2-knots.}


\Classification{16T25}

\renewcommand{\thefootnote}{\arabic{footnote}}
\setcounter{footnote}{0}

\section{Introduction}
\subsection{Yang--Baxter equation and its generalizations}

This work is mainly focusing on the {\em matrix} Yang--Baxter equation
\begin{gather}\label{YBE1}
R_{12}R_{13}R_{23}=R_{23}R_{13}R_{12} \in \operatorname{End}\big(V^{\otimes 3}\big),\qquad R\in \operatorname{End}\big(V^{\otimes 2}\big),
\end{gather}
and the {\em matrix} tetrahedral Zamolodchikov equation \cite{Zam}
\begin{gather}\label{TE1}
\Phi_{123}\Phi_{145}\Phi_{246}\Phi_{356}=\Phi_{356}\Phi_{246}\Phi_{145}\Phi_{123}\in \operatorname{End}\big(V^{\otimes 6}\big),\qquad \Phi\in \operatorname{End}\big(V^{\otimes 3}\big).
\end{gather}
In both cases $V$ is a f\/inite-dimensional vector space, the indices denote the numbers of the space copies in which linear operators act non-trivially.
We should also mention the universal description of the $n$-simplex equation (see, e.g.,~\cite{Hie+,Hie}), generalizing the Yang--Baxter and the Zamolodchikov equations.

The work is aimed to generalize one of the existing applications of the theory of the Yang--Baxter equation to the theory of quantum integrable systems. It should be noted that this equation and subsequently the theory of quantum groups marked a new era in the f\/ield of exactly-solvable models of mathematical physics. This concerned both models of statistical physics in dimension~2, and one-dimensional quantum mechanical models of the theory of magnets. The established connection turned out to be ef\/fective in the purely physical issues like the ef\/fect of the spontaneous magnetization as like as in many mathematical problems. The language of Hopf algebras not only expanded the machine of modern algebra but also allowed to explore new patterns, for example, in topology.

The mentioning of such diverse areas of modern mathematical physics here is not accidental. The transition from the Yang--Baxter equations to the tetrahedral equation appears natural from many points of view: low-dimensional topology, combinatorics, statistical models, topo\-lo\-gical f\/ield theories, homotopy algebraic structures. Here we present a brief table showing some heredity in subjects related to the Yang--Baxter equation and the tetrahedral one.

\begin{table}[h!]\centering
\caption{Relations.}\label{tab1}\vspace{1mm}
\begin{tabular}{|p{38mm}|p{51mm}|p{53mm}|}
\hline
 &Yang--Baxter equation & Zamolodchikov equation \\ \hline
statistical models & $d=2$ &$d=3$ \\ \hline
spin chains & $d=1$ & $d=2$ \\ \hline
homotopy Lie algebras & Lie algebras & $2$-Lie algebras (e.g., thesis by A.S.~Crans~\cite{Crans}) \\ \hline
topological invariants &Turaev--Reshetikhin-type knot invariants & $2$-knot quasi-invariants in~\cite{KST,KST+} \\ \hline
Hopf algebras & quasi-triangular Hopf algebras & ? \\ \hline
\end{tabular}
\end{table}

\subsection[Universal integrability in $d=1$]{Universal integrability in $\boldsymbol{d=1}$}
Besides the equation \eqref{YBE1}
\begin{gather*}
R_{12}R_{13}R_{23}=R_{23}R_{13}R_{12}
\end{gather*}
we consider the structural equations in the combinatorial form
\begin{gather}
R'_{12} R'_{23} R'_{12}=R'_{23} R'_{12} R'_{23}, \label{YBE}\\
R' L\otimes L=L\otimes L R', \label{RLL}
\end{gather}
where $R'\in \operatorname{End}(V)^{\otimes 2}$ and $L\in \operatorname{End}(V)\otimes \operatorname{End}(V_q)$, here $V_q$ is the quantum vector space (i.e., one another vector space, distinguished from~$V$.) The transition from \eqref{YBE1} to \eqref{YBE} can be realized as follows: if $R$ satisf\/ies~\eqref{YBE1} then $R'_{12}=R_{12}P_{12}$ as like as $R''_{12}=P_{12}R_{12}$ satisfy~\eqref{YBE}, here~$P_{12}$ is the transposition linear operator in $V\otimes V$. The equation~\eqref{RLL} is traditionally called the RLL-relation, it plays a substantial role in Faddeev--Reshetikhin--Takhtadzhyan algebras~\cite{FRT}. A~solution example of \eqref{RLL} is provided by
\begin{gather*}
L=R_{1i_1}R_{1i_2}\cdots R_{1i_k},
\end{gather*}
where $L$ is an operator in the quantum space $V_q=V_{i_1}\otimes\cdots \otimes V_{i_k}$, here $V_{i_j}$ is just a copy of the space~$V$.

The work \cite{Maillet} presents a construction of a commutative family in the algebra $\operatorname{End}(V_q)$ containing the trace $I_1=\operatorname{Tr}_V L$.
\begin{Lemma}[\cite{Maillet}]\label{Maillet_lemma}
Let us introduce a notation $L_i$ for the corresponding element in $\operatorname{End}(V_i)\otimes \operatorname{End}(V_q)$. Then the operators
\begin{gather}\label{Maillet_I}
I_k=\operatorname{Tr}_{1\cdots k}L_{1}\cdots L_k R'_{12}R'_{23}\cdots R'_{k-1,k}
\end{gather}
commute in $\operatorname{End}(V_q)$. The trace is meant with respect to the auxiliary spaces $V_i$.
\end{Lemma}

This statement has an important role in the technique of constructing quantum-mechanical integrable systems. This is applicable in quantum Gaudin systems, Ruijenaars--Schneider system and others. Moreover, this is directly associated with the theory of exactly-solvable models of statistical physics on $2$-dimensional lattices. The study of the spectrum of the transfer-matrix is a key ingredient in the problem of f\/inding the partition function asymptotics of some statistical models (e.g.,~\cite{baxter}).

\subsection{Tetrahedral equation}\label{tetra}
As like as in the Yang--Baxter case for many purposes it is more convenient to consider the \textit{set-theoretic tetrahedral equation} (STTE) which may be def\/ined as follows: let $X$ be a f\/inite set, we say that there is a solution for STTE on~$X$ if there is a map
\begin{gather*}
X\times X\times X\stackrel{\Phi}{\longrightarrow} X\times X\times X,
\end{gather*}
satisfying the relation (graphically coinciding with \eqref{TE1})
\begin{gather*}
\Phi_{123}\circ \Phi_{145}\circ \Phi_{246}\circ \Phi_{356} = \Phi_{356}\circ \Phi_{246}\circ \Phi_{145}\circ \Phi_{123}\colon \ X^{\times 6}\to X^{\times 6}.
\end{gather*}
Here, however, unlike \eqref{TE1}, $X^{\times 6}$ denotes the Cartesian 6-th power of $X$ and the subscripts denote the number of factors to which $\Phi$ is applied, in other factors the map acts identically. For example
\begin{gather*}
\Phi_{356}(a_1,a_2,a_3,a_4,a_5,a_6)=(a_1,a_2,\Phi_1(a_3,a_5,a_6),a_4,\Phi_2(a_3,a_5,a_6),\Phi_3(a_3,a_5,a_6))\\
\hphantom{\Phi_{356}(a_1,a_2,a_3,a_4,a_5,a_6)}{} =(a_1,a_2,a_3',a_4,a_5',a_6'),
\end{gather*}
where
\begin{gather*}
\Phi(x,y,z)=(\Phi_1(x,y,z),\Phi_2(x,y,z),\Phi_3(x,y,z))=(x',y',z').
\end{gather*}
One distinguishes the functional tetrahedral equation (FTE), for example the so-called electric solution \cite{KKS,KKS+}, represented as a transformation acting on the space of functions of three variables
\begin{gather*}
\Phi(x,y,z)=(x_1,y_1,z_1),\\ x_1=\frac {x y} {x+z+x y z},\qquad
y_1=x+z+x y z,\qquad
z_1=\frac{ yz} {x+z+x y z}.
\end{gather*}
This solution is relevant to the well-known star-triangle relation in the theory of electric circuits.

\begin{figure}[h!]
\centering
\includegraphics[width=9.8cm]{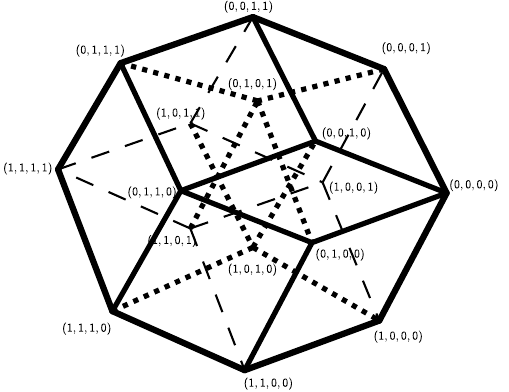}
\caption{Tesseract.}\label{tesseract}
\end{figure}

\begin{figure}[h!]
\centering
\includegraphics[width=6.7cm]{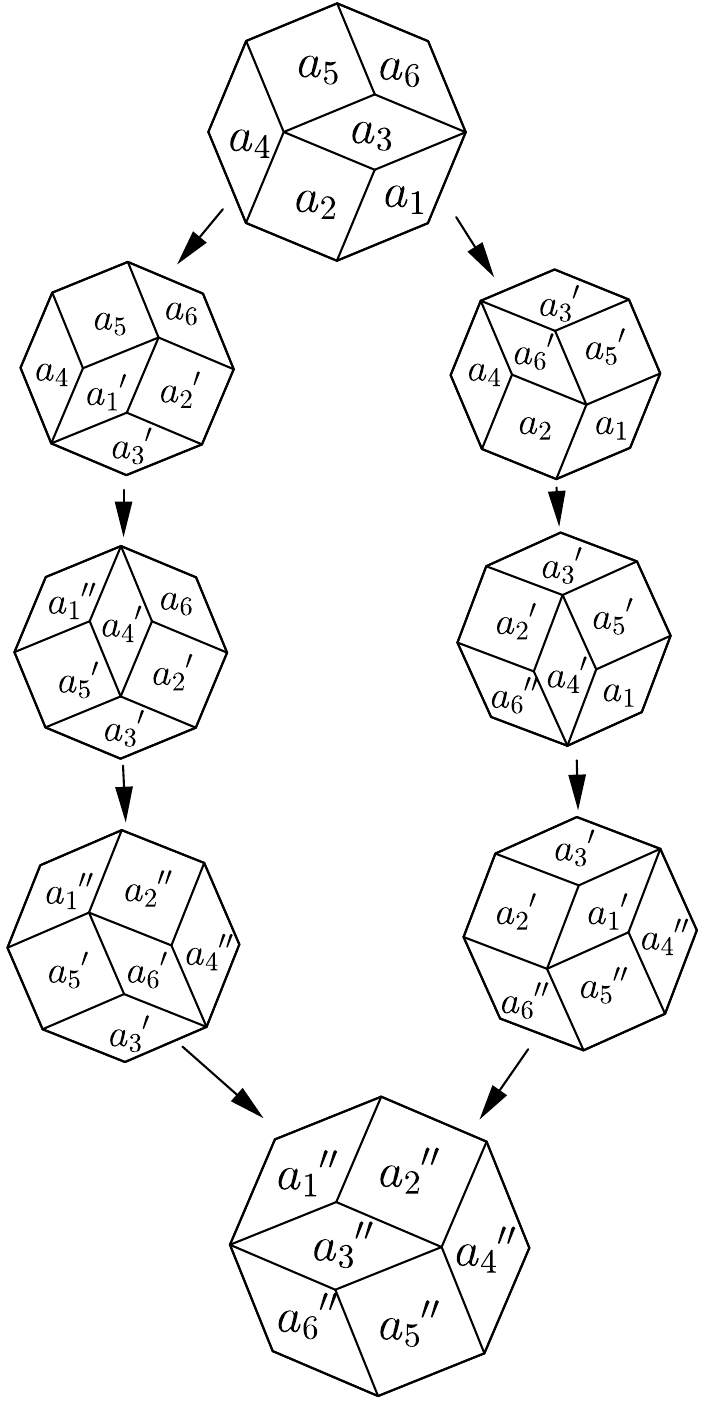}
\caption{Coloring.}\label{proc}
\end{figure}

There is another one interpretation of the tetrahedral equation in the task of coloring of $2$-faces of the $4$-dimensional cube with elements of the set $X$. Let $\Phi\colon X\times X\times X \rightarrow X \times X \times X$ be a map. A coloring is called admissible if the colors of the incoming faces of all $3$-cubes~$x$,~$y$,~$z$ are related with the colors of outgoing faces $x'$, $y'$, $z'$ by the action of the map
\begin{gather*}
\Phi\colon \ (x',y',z')=\Phi(x,y,z).
\end{gather*}
At Fig.~\ref{tesseract} we design a projection of the $4$-cube to a $3$-space
and at Fig.~\ref{proc} -- two alternative sequences of coloring steps.
It turns out that the condition of equivalence of the coloring obtained by these two ways is equivalent to the tetrahedral equation on~$\Phi$. The problem is described in more details in \cite{KST}.

The $N$-cube $2$-faces coloring problem allows us to construct a complex analogous to those calculating the Yang--Baxter cohomology for the case of set-theoretic tetrahedral equations in~\mbox{\cite{KST}}. The $3$-cocycles of the complex play a special role in this subject, they are determined by the condition
 \begin{gather*}
\varphi(a_1,a_2,a_3)\varphi(a_1',a_4,a_5)\varphi(a_2',a_4',a_6)\varphi(a_3',a_5',a_6') \\
 \qquad{} =\varphi(a_3,a_5,a_6)\varphi(a_2,a_4,a_6')\varphi(a_1,a_4',a_5')\varphi(a_1',a_2',a_3')
\end{gather*}
in the notation of Fig.~\ref{proc}. In particular the following lemma fulf\/ills:
\begin{Lemma}[{\cite[Theorem~4]{KST}}]\label{fut-vect}
Let $\Phi$ be a solution for the STTE, $\varphi$ be a $3$-cocycle of the tetrahedral complex. Let $V$ be the vector space generated by the elements of the set $X$. It has basis $\{e_x\colon x\in X\}$ marked by elements of $X$. Then let us define a linear operator~$A$ on~$V^{\otimes 3}$ specifying its values on tensor products of basis vectors. We say that
\begin{gather*}
A(s)(e_x\otimes e_y \otimes e_z)=\varphi(x,y,z)^s(e_{x'}\otimes e_{y'}\otimes e_{z'})\end{gather*}
if and only if $\Phi(x,y,z)=(x',y',z')$. In this case $A(s)$ provides a solution for the matrix tetrahedral equation.
\end{Lemma}
\begin{Remark}
To construct such matrix solution from the electric solution for the FTE one should realize the f\/inite-dimensional reduction of the latter. In~\cite[Lemma~6]{KST} it is demonstrated that the electric solution reduces correctly to the set $X$ obtained as follows: consider the residue ring~$\mathbb{Z}/p^k\mathbb{Z}$, where $p$ is either a prime number of the form $p=4l+1$, or $p=2$, and~$k$ is an integer $\ge 2$ (the Legendre symbol $\big({-}\frac{1}{p}\big)$ equals~$1$). We f\/ix one square root of $-1$ and call it $\varepsilon\in \mathbb{Z}/p\mathbb{Z}$. Our set $X$ will be the following subset of~$\mathbb{Z}/p^k\mathbb{Z}$
\begin{gather*}
X=\big\{x \in \mathbb{Z}/p^k\mathbb{Z} \colon x=\varepsilon \; \text{mod}\; p\big\}.
\end{gather*}
\end{Remark}

\section{Commutative family}

\subsection[The commutativity demonstration in $d=1$]{The commutativity demonstration in $\boldsymbol{d=1}$}
Let us present here our own proof of the Maillet result in Lemma~\ref{Maillet_lemma}. It demonstrates the main technique and suggests the path for generalization. In this section we denote by $R$ and $L$ solutions for the equations~\eqref{YBE} and~\eqref{RLL} (do not confuse with equation~\eqref{YBE1}). We assume that $R$ is invertible. We present here the Maillet generators~\eqref{Maillet_I} in a slightly more algebraic way
\begin{gather*}
I_k=\operatorname{Tr}_{V_1\otimes\cdots \otimes V_k} L^{\otimes k}R_{12}\cdots R_{k-1,k},
\end{gather*}
where the tensor product is taken with respect to the auxiliary spaces $V_i$, the product in quantum space $V_q$ is implied. To prove the commutativity of $I_k$ we demonstrate that there exists a linear operator $A\in \operatorname{End}(V)^{\otimes k+l}$ such that
\begin{gather*}
A L^{\otimes k+l}R_{12}\cdots R_{k-1,k}R_{k+1,k+2}\cdots R_{k+l-1,k+l}A^{-1}\\
\qquad{} = L^{\otimes k+l}R_{12}\cdots R_{l-1,l}R_{l+1,l+2}\cdots R_{k+l-1,k+l}.
\end{gather*}
This yields that the traces of these expressions with respect to the auxiliary spaces coincide. This in turn produce the identity
\begin{gather*}
[I_k,I_l]=0.
\end{gather*}
In what follows $\operatorname{Ad}_g X$ means the group adjoint action $gXg^{-1}$. Let us introduce some accessory notations: $R_{\overline{lm}}=R_{l,l+1}R_{l+1,l+2}\cdots R_{m-1,m}$ for $l<m$. This expression is subject to some relations
\begin{Lemma}\label{lem3}
\begin{gather*}
\operatorname{Ad}_{R_{\overline{1k}}}R_{m,m+1}=R_{m+1,m+2},\qquad 1\le m \le k-2.
\end{gather*}
\end{Lemma}

\begin{proof}
\begin{gather*}
R_{12}\cdots R_{m-1,m}R_{m,m+1}R_{m+1,m+2}\cdots R_{k-1,k} R_{m,m+1}R_{k-1,k}^{-1}\cdots\\
\qquad\quad{}\times R_{m+1,m+2}^{-1}R_{m,m+1}^{-1}R_{m-1,m}^{-1}\cdots R_{12}^{-1} \\
\qquad{} =R_{12}\cdots R_{m-1,m}R_{m,m+1}\underline{R_{m+1,m+2}R_{m,m+1} R_{m+1,m+2}^{-1}}R_{m,m+1}^{-1}R_{m-1,m}^{-1}\cdots R_{12}^{-1} \\
\qquad{}=R_{12}\cdots R_{m-1,m}\underline{R_{m,m+1}R_{m,m+1}^{-1}}R_{m+1,m+2} \underline{R_{m,m+1}R_{m,m+1}^{-1}}R_{m-1,m}^{-1}\cdots R_{12}^{-1} \\
\qquad{}=R_{12}\cdots R_{m-1,m}R_{m+1,m+2} R_{m-1,m}^{-1}\cdots R_{12}^{-1}=R_{m+1,m+2} . \tag*{\qed}
\end{gather*}\renewcommand{\qed}{}
\end{proof}

Another relation is expressed by the
\begin{Lemma}\label{lem4}
\begin{gather*}
\operatorname{Ad}_{R_{\overline{2k}}}\operatorname{Ad}_{R_{\overline{1,k-1}}} R_{k-1,k}=R_{12}.
\end{gather*}
\end{Lemma}

\begin{proof}We demonstrate the statement by induction. For $k=3$ we have
\begin{gather*}
R_{23}\underline{R_{12}R_{23}R_{12}^{-1}}R_{23}^{-1}=R_{23}\underline{R_{23}^{-1}R_{12}R_{23}}R_{23}^{-1}=R_{12}.
\end{gather*}
Let the statement be true for $k-1$, then
\begin{gather*}
\operatorname{Ad}_{R_{\overline{2k}}}\operatorname{Ad}_{R_{\overline{1,k-1}}} R_{k-1,k}=R_{23}\cdots R_{k-1,k}R_{12}\cdots R_{k-3,k-2}\underline{R_{k-2,k-1}R_{k-1,k}R_{k-2,k-1}^{-1}}\\
\qquad\quad{} \times R_{k-3,k-2}^{-1}\cdots R_{12}^{-1}R_{k-1,k}^{-1}\cdots R_{23}^{-1} \\
\qquad{}=R_{23}\cdots \underline{R_{k-1,k}}R_{12}\cdots R_{k-3,k-2} \underline{R_{k-1,k}^{-1}}R_{k-2,k-1}\underline{R_{k-1,k}}R_{k-3,k-2}^{-1}\cdots R_{12}^{-1}\underline{R_{k-1,k}^{-1}}\cdots R_{23}^{-1} \!\\
\qquad{} =R_{23}\cdots R_{k-2,k-1} R_{12}\cdots R_{k-3,k-2}R_{k-2,k-1}R_{k-3,k-2}^{-1}\cdots R_{12}^{-1}R_{k-2,k-1}^{-1}\cdots R_{23}^{-1} \\
\qquad{} =\operatorname{Ad}_{R_{\overline{2,k-1}}}\operatorname{Ad}_{R_{\overline{1,k-2}}} R_{k-2,k-1}=R_{12}.\tag*{\qed}
\end{gather*}\renewcommand{\qed}{}
\end{proof}

We may now fabricate an operator $A$ by the formula
\begin{gather*}
A=R_{\overline{l,k+l}} R_{\overline{l-1,k+l-1}}\cdots R_{\overline{1,k+1}}.
\end{gather*}

\begin{Lemma}
\begin{gather*}
\operatorname{Ad}_{A} (L^{\otimes k+l}R_{12}\cdots R_{k-1,k}R_{k+1,k+2}\cdots R_{k+l-1,k+l})\\
\qquad{}= L^{\otimes k+l}R_{12}\cdots R_{l-1,l}R_{l+1,l+2}\cdots R_{k+l-1,k+l}.
\end{gather*}
\end{Lemma}

\begin{proof}
Let us note that in virtue of \eqref{RLL} we need to demonstrate only
\begin{gather*}
\operatorname{Ad}_{A} (R_{12}\cdots R_{k-1,k}R_{k+1,k+2}\cdots R_{k+l-1,k+l})= R_{12}\cdots R_{l-1,l}R_{l+1,l+2}\cdots R_{k+l-1,k+l}.
\end{gather*}
On the other hand Lemma \ref{lem3} yields
\begin{gather}\label{eq11}
\operatorname{Ad}_{A} (R_{12}\cdots R_{k-1,k})=R_{l+1,l+2}\cdots R_{k+l-1,k+l}.
\end{gather}
Lemma \ref{lem4} provides
\begin{gather}\label{eq2}
\operatorname{Ad}_A(R_{k+1,k+2}\cdots R_{k+l-1,k+l})=R_{12} R_{23}\cdots R_{l-1,l}.
\end{gather}
The commutativity argument of the right hand sides of~\eqref{eq11} and~\eqref{eq2} f\/inishes the proof.
\end{proof}

This reasoning can be generalized for generating functions of integrals. Let us introduce a notation
\begin{gather*}
S(t)=L^{\otimes N}\big(1+tR_{12}+t^2R_{12}R_{23}+\cdots+t^{N-1}R_{12}\cdots R_{N-1,N}\big).
\end{gather*}
Then
\begin{gather*}
Q(t)=\operatorname{Tr} S(t)=\sum_{k=0}^{N-1}t^kI_kI_1^{N-k}.
\end{gather*}
If one now considers the conjugation operator
\begin{gather*}
A=R_{\overline{N,2N}} R_{\overline{N-1,2N-1}}\cdots R_{\overline{1,N+1}},
\end{gather*}
then
\begin{gather*}
\operatorname{Ad}_A(S(t)\otimes S(u))=S(u)\otimes S(t).
\end{gather*}
This immediately implies
\begin{gather*}
[Q(t),Q(u)]=0.
\end{gather*}

\subsection[Regular $3d$ lattices and statistical models]{Regular $\boldsymbol{3d}$ lattices and statistical models}

Consider a periodic three-dimensional lattice of size $K\times L\times M$, we mark the edges incoming to the node $(i,j,k)$ as $x_{i,j,k}$, $y_{i,j,k}$, $z_{i,j,k}$. The periodicity conditions imply $*_{K+1,j,k}=*_{1,j,k}$, and similar identities for other indexes. Consider a statistical model with the Boltzmann weights in the nodes of the lattice sites determined by the value of the 3-cocycle $\varphi$ of the tetrahedral complex. The states are def\/ined as admissible colorings of the edges, i.e., such that in each node the condition fulf\/ills
\begin{gather*}
\Phi(x_{i,j,k},y_{i,j,k},z_{i,j,k})=(x_{i+1,j,k},y_{i,j+1,k},z_{i,j,k+1}).
\end{gather*}
A partition function is def\/ined as follows
\begin{gather*}
Z(s)=\sum_{\text{col}}\prod_{i,j,k}\varphi(x_{i,j,k},y_{i,j,k},z_{i,j,k})^s.
\end{gather*}
To explore the ``integrability" of the subsidiary quantum problem one needs to recognize the layer-to-layer transfer-matrix. In order to determine what it is, we need another interpretation of the partition function.

We associate a copy of the space $V$ to each line of the lattice. For convenience, we denote the vertical spaces by characters $V_{ik}$ and the horizontal ones~-- by $E_i$ and $N_k$. We construct an operator $A_{ik}(s)$ acting in $E_i\otimes V_{ik}\otimes N_k$ with the chosen $3$-cocycle according to Lemma~\ref{fut-vect} pattern.
\begin{figure}[t]\centering
\includegraphics[width=120mm]{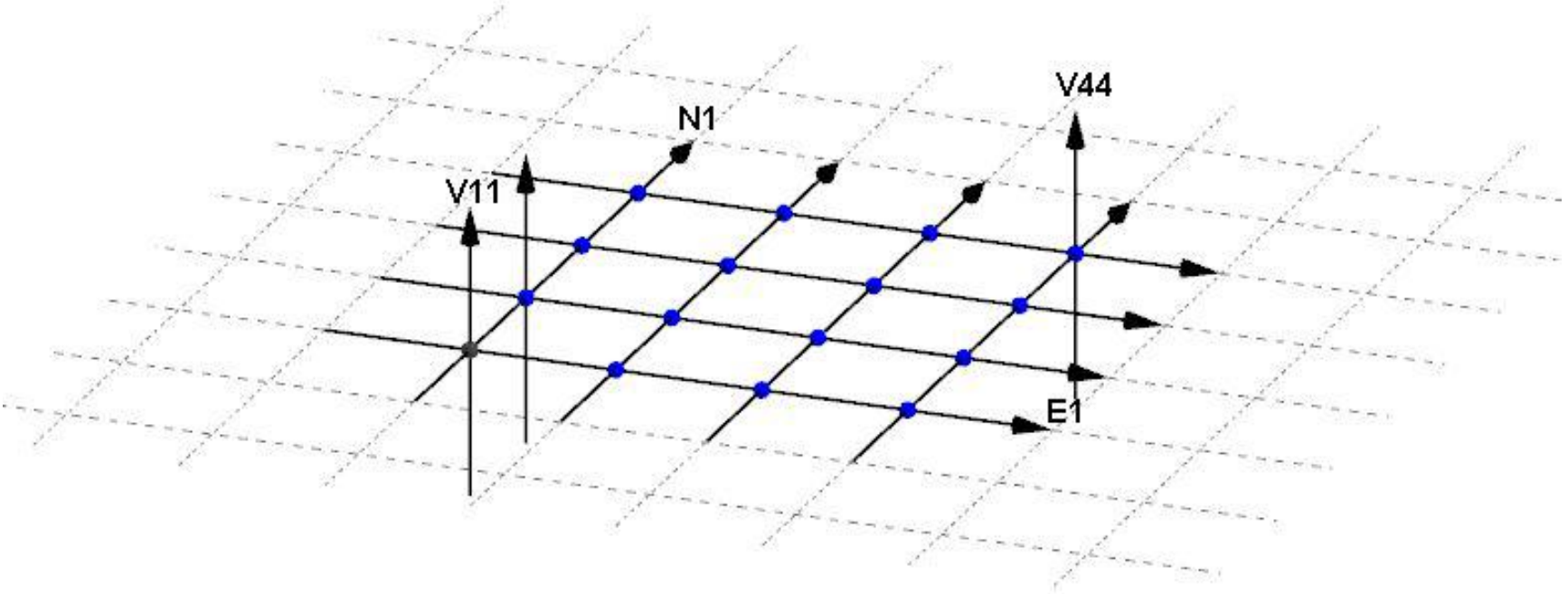}
\caption{1-layer conf\/iguration.}
\end{figure}
Let us def\/ine the transfer-matrix by a $1$-layer product
\begin{gather*}
T(s)=\operatorname{Tr}\prod_{\overrightarrow{i}}\prod_{\overrightarrow{k}} A_{ik}(s)=\operatorname{Tr}(B_1(s)\cdots B_K(s)),
\end{gather*}
where
\begin{gather*}
B_i(s)=A_{i1}(s)\cdots A_{iM}(s).
\end{gather*}
The formula implies the trace of matrices with respect to horizontal spaces. This operator acts on the tensor product of vertical spaces. It turns out that the partition function takes the form
\begin{gather*}
Z(s)=\operatorname{Tr}_{V_{jk}} T(s)^L.
\end{gather*}

Such issues as the asymptotic behavior of partition functions with respect to increasing the size of the lattice may be solved by the study of the spectrum of the transfer-matrix. The integrability condition, i.e., the possibility of including the transfer-matrix in a large commutative family, simplif\/ies the problem of f\/inding the spectrum.

\subsection{Some consequences of the tetrahedral equation}
In this section we give a generalization of the Maillet construction in the case of the three-dimensional lattice and the transfer-matrix associated with the solution of the matrix tetrahedral equation~$\Phi$. Consider a lattice of size $K\times L\times M$ and several forms of $1$-layer product
\begin{gather}
\Phi_{(i)*(j)}=\Phi_{(i_1\ldots i_K)*(j_1\dots j_M)}=\prod_{s\in\overrightarrow{(1,\ldots,K)}}^{t\in\overrightarrow{(1,\dots,M)}} \Phi_{i_s l^s_{t} j_t}\nonumber\\
\hphantom{\Phi_{(i)*(j)}}{} =\prod_{s\in\overrightarrow{(1,\dots,K)}}\Phi_{i_s(l^s_{1}\dots l^s_{M})(j_1\dots j_M)}=\prod_{s\in\overrightarrow{(1,\dots,K)}}\Phi_{i_s(l^{s})(j)}.\label{eq1}
\end{gather}
The arrows over indexing sets mean the direction in products: for example
\begin{gather*}
\prod_{s\in\overrightarrow{(1,\ldots,K)}} A_s=A_1\cdots A_K.
\end{gather*}
In the last two parts of \eqref{eq1} we use notations
\begin{gather*}
\Phi_{k(i)(j)}=\Phi_{k(i_1\ldots i_t)(j_1\ldots j_t)}=\prod_{s\in\overrightarrow{(1,\ldots,t)}}\Phi_{k i_s j_s},\\
\Phi_{(i)(j)k}=\Phi_{(i_1\ldots i_t)(j_1\ldots j_t)k}=\prod_{s\in\overrightarrow{(1,\ldots,t)}}\Phi_{i_s j_s k},
\end{gather*}
where dif\/ferent letters in indices correspond to dif\/ferent copies of the space $V$.

\begin{Remark}
Let us make some comments on the indices rule. Here $(i)$ is a multiindex $(i_1,\ldots,i_s)$. In what follows we use sequences of multiindices, we denote by superscript the number of a~multiindex, for example by $(i^l)$ we denote the multiindex $(i_1^l,\ldots,i_s^l)$ remarking each time the dimension~$s$. The dimension depends on the lattice parameters~$K$,~$L$. The construction at all does, but for the readability reasons we omit this parameters in the def\/inition of the commutative family and point out the values for each instance explicitly.
\end{Remark}

The transfer-matrix can be represented as the trace of the expression \eqref{eq1}
\begin{gather*}
T=I_1=\operatorname{Tr}_{(i)(j)}\Phi_{(i)*(j)},
\end{gather*}
here $(i)$ is a $K$-multiindex, $(j)$ is an $M$-one. Let us write down several identities which are straightforward consequences of the tetrahedral equation. They can be considered as genera\-li\-zations of the $RLL$-relations.
\begin{Lemma}\label{lem_RLL}\sloppy
\begin{gather}
\Phi_{123}\Phi_{1(i)(j)}\Phi_{2(i)(l)}\Phi_{3(j)(l)}=\Phi_{3(j)(l)}\Phi_{2(i)(l)}\Phi_{1(i)(j)}\Phi_{123},\label{cons}\\
\Phi_{(i)(j)1}\Phi_{(i)(l)2}\Phi_{(j)(l)3}\Phi_{123}=\Phi_{123}\Phi_{(j)(l)3}\Phi_{(i)(l)2}\Phi_{(i)(j)1}, \label{cons2}\\
\Phi_{(i)(i')0}\Phi_{(i)*(j)}\Phi_{(i')*(j')}\Phi_{0(j)(j')}=\Phi_{0(j)(j')}\Phi_{(i')*(j')}\Phi_{(i)*(j)}\Phi_{(i)(i')0},
\label{cons3}
\end{gather}
for $s$-multiindices $(i)$, $(j)$, $(l)$ in \eqref{cons}, \eqref{cons2} and $K$-multiindices $(i)$, $(i')$ and $M$-multi\-in\-di\-ces~$(j)$,~$(j')$ in~\eqref{cons3}.
\end{Lemma}
The proof is placed in Appendix~\ref{app1}.

We introduce also some twisted versions of solutions for the tetrahedral equation
\begin{gather*}
\Phi_{123}^L=P_{12}\Phi_{123},\qquad \tilde{\Phi}_{123}^L=P_{23}\Phi_{123},\qquad \Phi_{123}^R=\Phi_{123}P_{23},\qquad \tilde{\Phi}_{123}^R=\Phi_{123}P_{12}.
\end{gather*}
The identities fulf\/ill
\begin{gather}
\Phi_{123}^L \Phi_{145}\Phi_{246} \Phi_{356}^R= \Phi_{356}^R\Phi_{145}\Phi_{246}\Phi_{123}^L,\label{L--R}\\
\tilde{\Phi}^L_{\alpha\beta\gamma}\Phi_{\alpha 12}^R\Phi_{\beta 23} \Phi_{\gamma 13}=\Phi_{\beta 23}\Phi_{\gamma 13}
\Phi_{\alpha 12}^R\tilde{\Phi}^L_{\alpha \beta \gamma}, \label{LR--}\\
\Phi_{145}^R\Phi_{123}^R\Phi_{356}^L\Phi_{246}^L= \Phi_{356}^L\Phi_{246}^L\Phi_{145}^R\Phi_{123}^R, \label{RRLL}\\
\tilde{\Phi}_{123}^L\tilde{\Phi}_{145}^L\tilde{\Phi}_{246}^R\tilde{\Phi}_{356}^R=
\tilde{\Phi}_{246}^R\tilde{\Phi}_{356}^R\tilde{\Phi}_{123}^L\tilde{\Phi}_{145}^L.\label{tRRLL}
\end{gather}
They perform a role similar to the Yang--Baxter equation in combinatorial notation. These equalities are also simple consequences of the tetrahedral equation, we do not provide here the proofs with the only purpose to simplify the exposition.

\subsection{Two families}
In this section we always consider the slice of the regular cubic lattice with $K\times L\times M$ vertices. The $2$-dimensional layer is def\/ined by f\/ixing the $2$-nd coordinate and has the dimension~$K\times M$. The main constructing block of what follows is~$\Phi_{(i)*(j)}$ def\/ined by~\eqref{eq1}.

First we pay attention to the connection of the tetrahedral equation and the Yang--Baxter equations of the following type: for an invertible $\Phi$ the formula~\eqref{cons} can be transformed to the kind
\begin{gather*}
 \Phi_{123}\Phi_{1(i)(j)}\Phi_{2(i)(l)}\Phi_{3(j)(l)}\Phi_{123}^{-1} =\Phi_{3(j)(l)}\Phi_{2(i)(l)}\Phi_{1(i)(j)},
\end{gather*}
where $(i)$, $(j)$ and $(l)$ are multiindices of the same dimension~$s$. Taking the trace of both parts with respect to indices $1$, $2$, $3$, one deduces that the expression $R_{(i)(j)}=\operatorname{Tr}_{1}\Phi_{1(i)(j)}$ satisf\/ies the Yang--Baxter equation \eqref{YBE1} in the tensor product $V^{\otimes s}$. Moreover as a consequence of~\eqref{cons3} we have
\begin{gather*}
\Phi_{(i)(i')0}\Phi_{(i)*(j)}\Phi_{(i')*(j')}\Phi_{0(j)(j')}\Phi^{-1}_{(i)(i')0}=\Phi_{0(j)(j')}\Phi_{(i')*(j')}\Phi_{(i)*(j)},
\end{gather*}
here $(i)$ and $(i')$ are of dimension $K$, $(j)$, $(j')$ are of dimension~$M$. If one takes the trace of both parts with respect to the spaces~$(i)$, $(i')$, $0$ it turns out that the expression $L_{*(j)}=\operatorname{Tr}_{(i)}\Phi_{(i)*(j)}$ satisf\/ies the $RLL$ relation
\begin{gather*}
L_{*(j)}L_{*(j')}R_{(j)(j')}=R_{(j)(j')}L_{*(j')}L_{*(j)}.
\end{gather*}
Thus the data
\begin{gather*}
L_{*(j)}=\operatorname{Tr}_{(i)}\Phi_{(i)*(j)},\qquad R'_{(i)(j)}=\operatorname{Tr}_{1}\Phi^R_{1(i)(j)}
\end{gather*}
meets Lemma~\ref{Maillet_lemma} conditions. The immediate consequence of this is that the following set of operators
\begin{gather*}
I_{0,k}=\operatorname{Tr}_{(j^l),l=1,\ldots,k}\prod_{l=\overrightarrow{1,\ldots,k}}L_{*(j^l)}\prod_{m=\overrightarrow{1,\ldots,k-1}}R'_{(j^m)(j^{m+1})}
\end{gather*}
is commutative and contains $I_1=\operatorname{Tr}_{(j)} L_{*(j)}=\operatorname{Tr}_{(i)(j)}\Phi_{(i)*(j)}$. Here multiindices $(j^m)$ have dimension~$M$. There is a version of the formula for $I_{0,k}$
\begin{gather*}
I_{0,k}=\operatorname{Tr}_{(i^l)(j^l)(s)}\prod_{l=\overrightarrow{1,\ldots,k}}\Phi_{(i^l)*(j^l)}\prod_{m=\overrightarrow{1,\ldots,k-1}}\Phi^R_{s_m (j^m)(j^{m+1})}
\end{gather*}
with multiindices $(i^l)$ of dimension $K$, $(j^l)$ of dimension $M$ and $(s)$ of dimension $k-1$. A similar argument can show that, due to Lemma~\ref{Maillet_lemma} the family
\begin{gather}
I_{n,0}=\operatorname{Tr}_{(i^l)(j^l)(t)}\prod_{l=\overrightarrow{1,\ldots,n}}\Phi_{(i^l)*(j^l)}\prod_{m=\overrightarrow{1,\ldots,n-1}}\Phi^L_{(i^m)(i^{m+1})t_m}\nonumber
\end{gather}
is also commutative and includes $I_1$, here the multiindices have the same dimension except $(m)$ which is $n-1$. The main accomplishment of the paper is that these two families commute among themselves. In order to give a precise formulation let us introduce some notation:
\begin{gather*}
B_k=\Phi_{(i^1)\ast(j^1)}\cdots\Phi_{(i^k)\ast(j^k)}=\prod_{\alpha\in\overrightarrow{(1,\ldots,k)}}\Phi_{(i^\alpha)\ast(j^\alpha)},\\
R^s_k=\Phi_{s_1(j^1)(j^2)}^R\cdots\Phi_{s^{k-1}(j^{k-1})(j^k)}^R=\prod_{\alpha\in\overrightarrow{(1,\ldots,k-1)}}\Phi^R_{s_\alpha(j^\alpha)(j^{\alpha+1})},\\
L^s_l=\Phi_{(i^1)(i^2)s_1}^L\cdots\Phi_{(i^{l-1})(i^l)s_{l-1}}^L=\prod_{\alpha\in\overrightarrow{(1,\ldots,l-1)}}\Phi^L_{(i^\alpha)(i^{\alpha+1})s_\alpha},\\
A^p_L=\prod_{\overrightarrow{\alpha}\overleftarrow{\beta}}\Phi_{(i^\alpha)(i^\beta)p_{\alpha\beta}},\qquad
A^p_R=\prod_{\overrightarrow{\beta}\overleftarrow{\alpha}}\Phi_{p_{\alpha\beta}(j^\alpha)(j^\beta)}.
\end{gather*}
Here $(i^\alpha)$ is a $K$-multiindex, $(j^\beta)$ is an $M$-multiindex. $p_{\alpha\beta}$ is the set of accessory indices in last two formulas.
In addition, we need two auxiliary elements
\begin{gather*}
\Phi_p=\prod_{\overrightarrow{\alpha}\overleftarrow{\beta}}\tilde{\Phi}^L_{s_\alpha p_{\alpha+1,\beta}p_{\alpha,\beta}},\qquad
\Phi^*_p=\prod_{\overrightarrow{\beta}\overleftarrow{\alpha}}\tilde{\Phi}^R_{p_{\alpha,\beta+1}p_{\alpha,\beta}t_\beta}.
\end{gather*}

\subsection[The commutativity proof in $d=2$]{The commutativity proof in $\boldsymbol{d=2}$}
\begin{Definition}Let us call a solution of the tetrahedral equation generic if it is invertible and the operator $\operatorname{Tr}_{p}(A_R^p \Phi_p^*) $ is invertible too.
\end{Definition}
\begin{Theorem}
For any generic solution $\Phi$ of the tetrahedral equation the following is true
\begin{gather*}
[I_{0,k},I_{n,0}]=0.
\end{gather*}
\end{Theorem}

\begin{proof}
Let us consider an expression
\begin{gather*}
\underbrace{A_L^p}_{i^\alpha,i^\beta,p}\underbrace{\Phi_p}_{p,s} \underbrace{B_{k}}_{i^\alpha,j^\alpha}\otimes\underbrace{B_{l}}_{i^\beta,j^\beta}
\underbrace{R_k^s}_{j^\alpha,s} \underbrace{L_{l}^t}_{i^\beta,t}
\underbrace{A_R^p}_{j^\alpha,j^\beta,p} \underbrace{\Phi_p^*}_{p,t}.
\end{gather*}
The indices under each multiplier indicate the spaces in which it acts. Next, we shall omit the tensor product sign, considering the indices of the corresponding spaces. This expression can be transmuted to the form
\begin{gather*}
A_L^p B_{k}B_{l}\boxed{\Phi_p R_k^s A_R^p }L_l^t \Phi_p^*
\overset{\text{Lemma \ref{l1}}}{=} A_L^p B_{k}B_{l}\boxed{ A_R^p R_k^s \Phi_p }L_l^t \Phi_p^*\\
\qquad{}=\boxed{A_L^p B_{k} B_{l}A_R^p} R_k^s \Phi_p L_l^t \Phi_p^*
\overset{\text{Lemma \ref{l2}}}{=}\boxed{A_R^p B_{l} B_k A_L^p} R_k^s \Phi_p L_l^t \Phi_p^*\\
\qquad{} =A_R^p B_{l} B_k A_L^p R_k^s L_l^t\boxed{ \Phi_p \Phi_p^* }
\overset{\text{Lemma \ref{l3}}}{=}A_R^p B_{l} B_k A_L^p R_k^s L_l^t\boxed{ \Phi_p^* \Phi_p } \\
\qquad{} =A_R^p B_l B_k R_k^s\boxed{A_L^p L_l^t \Phi_p^* } \Phi_p
\overset{\text{Lemma \ref{l4}}}{=}A_R^p B_l B_k R_k^s \boxed{\Phi_p^* L_l^t A_L^p }\Phi_p \\
\qquad{} =A_R^p \Phi_p^* B_l B_k R_k^s L_l^t A_L^p \Phi_p.
\end{gather*}
Then we deduce
\begin{gather*}
{A_L^p}{\Phi_p}{B_{k} B_l}{R_k^s}{L_l^t}{A_R^p}{\Phi_p^*}({A_L^p}{\Phi_p})^{-1}=A_R^p \Phi_p^* B_l B_k R_k^s L_l^t.
\end{gather*}
In particular this implies
\begin{gather*}
\operatorname{Tr}_{(i^\alpha)(i^\beta)(p),s_\alpha}{B_{k+l}}{R_k^s}{L_l^t}{A_R^p}{\Phi_p^*}=
\operatorname{Tr}_{(i^\alpha)(i^\beta)(p)s_\alpha}A_R^p \Phi_p^* B_l B_k R_k^s L_l^t.
\end{gather*}
Let us note that the trace procedure factorizes
\begin{gather*}
\operatorname{Tr}_{(i^\alpha)(i^\beta)s_\alpha}\big({B_{k}B_l}{R_k^s}{L_l^t}\big)\operatorname{Tr}_{(p)}\big({A_R^p}{\Phi_p^*}\big)=
\operatorname{Tr}_{(p)}\big(A_R^p \Phi_p^*\big) \operatorname{Tr}_{(i^\alpha)(i^\beta) s_\alpha} \big( B_l B_k R_k^s L_l^t\big).
\end{gather*}
One may notice that
\begin{gather*}
\big(\operatorname{Tr}_{(p)} \big(A_R^p \Phi_p^* \big) \big)^{-1}\operatorname{Tr}_{(i^\alpha)(i^\beta) s_\alpha}\big({B_k B_l}{R_k^s}{L_l^t}\big)\operatorname{Tr}_{(p)}\big({A_R^p}{\Phi_p^*}\big)=
\operatorname{Tr}_{(i^\alpha)(i^\beta) s_\alpha} \big( B_l B_k R_k^s L_l^t\big).
\end{gather*}
Taking a trace over the remaining auxiliary spaces completes the proof
\begin{gather*}
I_{0k}I_{l0}=\operatorname{Tr}_{(j^\alpha)(j^\beta) t_\beta}\big(\operatorname{Tr}_{(i^\alpha)(i^\beta) s_\alpha}\big({B_{k+l}}{R_k^s}{L_l^t}\big)\big) \\
\hphantom{I_{0k}I_{l0}}{} =
\operatorname{Tr}_{(j^\alpha)(j^\beta) t_\beta}\big(\operatorname{Tr}_{(i^\alpha)(i^\beta) s_\alpha} \big( B_l B_k R_k^s L_l^t\big)\big)=I_{l0}I_{0k}.\tag*{\qed}
\end{gather*}\renewcommand{\qed}{}
\end{proof}

Proceed now to the proof of lemmas.
\begin{Lemma}\label{l1}
\begin{gather}\label{eq14}
\Phi_p R_k^s A_R^p=A_R^p R_k^s \Phi_p.
\end{gather}
\end{Lemma}

\begin{proof}
Let us deduce a useful formula
\begin{gather}\label{eq12}
\tilde\Phi^L_{\alpha\beta\gamma}\Phi^R_{\alpha (i)(j)}\Phi_{\beta(j)(k)}\Phi_{\gamma (i)(k)}=\Phi_{\beta(j)(k)}\Phi_{\gamma (i)(k)}\Phi^R_{\alpha (i)(j)}\tilde \Phi^L_{\alpha\beta\gamma}.
\end{gather}
It is a direct consequence of equation \eqref{LR--}. Let us write more thoroughly \eqref{eq14}
\begin{gather*}
\prod_{\overrightarrow{\alpha}\overleftarrow{\beta}}
\tilde{\Phi}^L_{s_\alpha p_{\alpha+1,\beta}p_{\alpha,\beta}}
\prod_{\overrightarrow{\alpha}}\Phi^R_{s_\alpha (j^\alpha)(j^{\alpha+1})}\prod_{\overrightarrow{\beta}\overleftarrow{\alpha}}\Phi_{p_{\alpha,\beta}(j^\alpha)(j^\beta)}.
\end{gather*}
Note that the left and right multipliers commute at dif\/ferent values of $\beta$ so it is enough to check the relation at f\/ixed $\beta=\beta_0$ performing further dressing starting with the multipliers closest to the center
\begin{gather*}
\begin{split}& \prod_{\overrightarrow{\alpha}}
\tilde{\Phi}^L_{s_\alpha p_{\alpha+1,\beta_0}p_{\alpha,\beta_0}}
\prod_{\overrightarrow{\alpha}}\Phi^R_{s_\alpha (j^\alpha)(j^{\alpha+1})}\prod_{\overleftarrow{\alpha}}\Phi_{p_{\alpha,\beta_0}(j^\alpha)(j^{\beta_0})} \\
& \qquad{} =\prod_{\overrightarrow{\alpha}}\big(\tilde{\Phi}^L_{s_\alpha p_{\alpha+1,\beta_0}p_{\alpha,\beta_0}}
\Phi^R_{s_\alpha (j^\alpha)(j^{\alpha+1})}\big)\prod_{\overleftarrow{\alpha}}\Phi_{p_{\alpha,\beta_0}(j^\alpha)(j^{\beta_0})}.
\end{split}
\end{gather*}
Now we will focus attention on how the elements of the left product are transferred through the right one. This can be done consecutively: f\/ix the index $\alpha=\alpha_0$ in the left product
\begin{gather*}
\tilde{\Phi}^L_{s_{\alpha_0} p_{\alpha_0+1,\beta_0}p_{\alpha_0,\beta_0}}
\Phi^R_{s_\alpha (j^{\alpha_0})(j^{\alpha_0+1})}\prod_{\overleftarrow{\alpha}}\Phi_{p_{\alpha,\beta_0}(j^\alpha)(j^{\beta_0})}.
\end{gather*}
The only multipliers of the right product which do not commute with the elements on the left have indices $\alpha=\alpha_0$ and $\alpha=\alpha_0+1$. When moving through them we use equality \eqref{eq12}
\begin{gather*}
\tilde{\Phi}^L_{s_{\alpha_0} p_{\alpha_0+1,\beta_0}p_{\alpha_0,\beta_0}}
\Phi^R_{s_\alpha (j^{\alpha_0})(j^{\alpha_0+1})}\Phi_{p_{\alpha_0+1,\beta_0}(j^{\alpha_0+1})(j^{\beta_0})}\Phi_{p_{\alpha_0,\beta_0}(j^{\alpha_0})(j^{\beta_0})} \\
\qquad{} =\Phi_{p_{\alpha_0+1,\beta_0}(j^{\alpha_0+1})(j^{\beta_0})}\Phi_{p_{\alpha_0,\beta_0}(j^{\alpha_0})(j^{\beta_0})}\Phi^R_{s_\alpha (j^{\alpha_0})(j^{\alpha_0+1})}\tilde{\Phi}^L_{s_{\alpha_0} p_{\alpha_0+1,\beta_0}p_{\alpha_0,\beta_0}}.\tag*{\qed}
\end{gather*}\renewcommand{\qed}{}
\end{proof}

\begin{Lemma}\label{l2}
\begin{gather*}
A_L^p B_{k} B_{l}A_R^p=A_R^p B_{l} B_k A_L^p.
\end{gather*}
\end{Lemma}

\begin{proof}
The statement is an immediate consequence of equation~\eqref{cons2}. We will illustrate the basic techniques of generalization. We write the left part of the expression
\begin{gather*}
\prod_{\overrightarrow{\alpha}\overleftarrow{\beta}}\Phi_{(i^\alpha)(i^\beta)p_{\alpha\beta}}\prod_{\overrightarrow{\alpha}}
\Phi_{(i^\alpha)\ast(j^\alpha)}
\prod_{\overrightarrow{\beta}}\Phi_{(i^\beta)\ast(j^\beta)}
\prod_{\overleftarrow{\alpha}\overrightarrow{\beta}}
\Phi_{p_{\alpha\beta}(j^\alpha)(j^\beta)}.
\end{gather*}
Here $(i^\alpha)$ and $(i^\beta)$ are $K$-multiindices, and $(j^\alpha)$ and $(j^\beta)$ are $M$-ones. We are going to move the multipliers of the third product through the multipliers of the second one. For doing this we use a twisting multipliers of the f\/irst and fourth factors. Note that the multipliers of the f\/irst and fourth factors commute except those with both coinciding indices. It remains to verify that the multiplier order is suitable. We will move the elements of the third product consequently from the left through the second product. To do this, the order relation on~$\beta$ in the f\/irst product should be such that the junior members stand on the right, and in the fourth on the left. Similarly, one can check that the order relation on $\alpha$ in the f\/irst product should be such that the senior members are on the right, and in the fourth on the left.
\end{proof}

\begin{Lemma}\label{l3}
\begin{gather*}
 \Phi_p \Phi_p^* = \Phi_p^* \Phi_p.
\end{gather*}
\end{Lemma}

\begin{proof}
Let us write the left side of the expression
\begin{gather*}
\prod_{\overrightarrow{\alpha}\overleftarrow{\beta}}\tilde{\Phi}^L_{s_\alpha p_{\alpha+1,\beta}p_{\alpha,\beta}}\prod_{\overrightarrow{\beta}\overleftarrow{\alpha}}
\tilde{\Phi}^R_{p_{\alpha,\beta+1}p_{\alpha,\beta}t_\beta}.
\end{gather*}
We move the right product through the left one component-wise, for f\/ixed $\alpha$ on the left and f\/ixed $\beta$ on the right. Let us consider these multipliers
\begin{gather*}
\prod_{\overleftarrow{\beta}}\tilde{\Phi}^L_{s_{\alpha_0} p_{\alpha_0+1,\beta}p_{\alpha_0,\beta}}\prod_{\overleftarrow{\alpha}}
\tilde{\Phi}^R_{p_{\alpha,\beta_0+1}p_{\alpha,\beta_0}t_{\beta_0}}.
\end{gather*}
The only non-commutative elements of these products have neighboring indices $\beta=\beta_0,\beta_0+1$ and $\alpha=\alpha_0,\alpha_0+1$. We write only them
\begin{gather*}
\tilde{\Phi}^L_{s_{\alpha_0} p_{\alpha_0+1,\beta_0+1}p_{\alpha_0,\beta_0+1}}
\tilde{\Phi}^L_{s_{\alpha_0} p_{\alpha_0+1,\beta_0}p_{\alpha_0,\beta_0}}
\tilde{\Phi}^R_{p_{\alpha_0+1,\beta_0+1}p_{\alpha_0+1,\beta_0}t_{\beta_0}}
\tilde{\Phi}^R_{p_{\alpha_0,\beta_0+1}p_{\alpha_0,\beta_0}t_{\beta_0}}.
\end{gather*}
Note that the right pair can be moved through the left pair of multipliers according to~\eqref{RRLL}.
\end{proof}

\begin{Lemma}\label{l4}
\begin{gather*}
A_L^p L_l^t \Phi_p^*=\Phi_p^* L_l^t A_L^p.
\end{gather*}
\end{Lemma}

\begin{proof}
Analogously to Lemma~\ref{l2}.
\end{proof}

\section{Conclusion}
The choice of the notations $I_{l,0}$ and $I_{0,k}$ used in the main part of the work pursued a def\/initive goal. We hope that the following hypothesis is correct
\begin{Hyp}
For any generic solution of the tetrahedral equation it can be constructed a~two-parameter commutative family $I_{l,k}$, which includes two families presented above.
\end{Hyp}

\begin{Remark}
Note that in the work \cite{BS} it is constructed a two-parameter family of operators commuting with a layer-by-layer transfer-matrix related to the special solution of the tetrahedral equation. It is curious to compare the result in the $q$-oscillator realization case and the result presented here.
\end{Remark}
However, the main result of this work reveals some important perspectives:
\begin{itemize}\itemsep=0pt
\item Primarily, it provides the opportunity to study the spectra of the corresponding quantum models and models of statistical physics in dimension $d=3$. We hope that in this case the Bethe ansatz technique also can be applied.
\item Another interesting direction is the generalization of the notion of integrability in the case of $2$-dimensional surfaces comprising the classical case. It is interesting to juxtapose the approach of this work with the language of paper~\cite{Z} on the theory of the Hitchin systems on surfaces. We would also be enthusiastic in the study of moduli spaces of 2-bundles on surfaces and an appropriate analogue of the Hitchin theory. In this case one has a~signif\/icant dif\/f\/iculty in constructing nonabelian gerbs.
\item
In \cite{KST,KST+} we present a construction of quasi-invariant of $2$-knots in the form of a partition function on the graph of double points of the $2$-knot diagram. This structure is resembling to the approach of \cite{JSC}. The similarity of the partition function expressions is intriguing and gives hope for further development activities in low-dimensional topology and topological quantum f\/ield theory. Presumably a certain connection of this method with four-dimensional quantum f\/ield theories, like the BF-theory, can be established.
\end{itemize}

\appendix

\section{Proof of Lemma \ref{lem_RLL}}\label{app1}
\begin{proof}
 Let us prove the formula \eqref{cons}. The rest is similar. The left hand side of \eqref{cons} can expressed as follows
\begin{gather*}
 \Phi_{123}\Phi_{1(i)(j)}\Phi_{2(i)(l)}\Phi_{3(j)(l)}=\Phi_{123}
 \prod_{s\in\overrightarrow{(1,\ldots,t)}}\Phi_{1 i_s j_s} \prod_{u\in\overrightarrow{(1,\ldots,t)}}\Phi_{2 i_u l_u} \prod_{v\in\overrightarrow{(1,\ldots,t)}}\Phi_{3 j_v l_v}.
 \end{gather*}
 The last product can be rearranged due to the commutativity of operators in dif\/ferent tensor components
 \begin{gather*}
 \Phi_{123}\prod_{s\in\overrightarrow{(1,\ldots,t)}}\Phi_{1 i_s j_s}\Phi_{2i_s l_s}\Phi_{3 j_s l_s}.
 \end{gather*}
 Then we apply the Zamoolodchkov equation with $\Phi_{123}$ to each factor consecutively from the left
 \begin{gather*}
 \prod_{s\in\overrightarrow{(1,\ldots,t)}}\Phi_{3 j_s l_s}\Phi_{2i_s l_s} \Phi_{1 i_s j_s}\Phi_{123}
 \end{gather*}
 and again rearrange the product
 \begin{gather*}
 \prod_{v\in\overrightarrow{(1,\ldots,t)}}\Phi_{3 j_v l_v} \prod_{u\in\overrightarrow{(1,\ldots,t)}}\Phi_{2 i_u l_u} \prod_{s\in\overrightarrow{(1,\ldots,t)}}\Phi_{1 i_s j_s}\Phi_{123} =\Phi_{3(j)(l)}\Phi_{2(i)(l)}\Phi_{1(i)(j)}\Phi_{123} .\tag*{\qed}
 \end{gather*}\renewcommand{\qed}{}
\end{proof}

\subsection*{Acknowledgements}
The work was partially supported by the RFBR grant of Russian Foundation of Basic Research 17-01-00366A, grant of the support of scientif\/ic schools 4833.2014.1, and the grant of the Dynasty Foundation. I would like to thank the referees for careful reading of the manuscript and constructive remarks on the material exposition.

\pdfbookmark[1]{References}{ref}
\LastPageEnding

\end{document}